\documentclass[useAMS,usenatbib,fleqn]{mn2e}
\usepackage{txfonts}
\usepackage[pdftex]{graphicx}
\usepackage{graphicx}
\usepackage{natbib}
\def\mpc{h^{-1} {\rm{Mpc}}}
\def\kms {\rm{km~s^{-1}}}
\def\apj {ApJ}
\def\apjl {ApJL}
\def\apjs {ApJS}
\def\aj {AJ}
\def\aap {A\&A}
\def\mnras {MNRAS}
\begin{document}
\title[Galaxies infalling into groups]
{Galaxies infalling into groups: filaments vs. isotropic infall}
\author[H.J. Mart\'inez et al.]
{H\'ector J. Mart\'inez\thanks{E-mail: julian@oac.unc.edu.ar} , 
Hern\'an Muriel and Valeria Coenda\\
Instituto de Astronom\'{\i}a Te\'orica y Experimental (IATE), 
CONICET$-$Universidad Nacional de 
C\'ordoba, Laprida 854, X5000BGR, C\'ordoba, Argentina\\
Observatorio Astron\'omico, Universidad Nacional de 
C\'ordoba, Laprida 854, X5000BGR, C\'ordoba, Argentina
}
\date{Accepted 2015 October 1. Received 2015 September 8; in original form 2015 July 16}
\pagerange{\pageref{firstpage}--\pageref{lastpage}} 
\maketitle
\label{firstpage}
\begin{abstract}
We perform a comparative analysis of the properties of galaxies infalling into
groups classifying them accordingly to whether they are: falling along 
filamentary structures; or they are falling isotropically. 
For this purpose, we identify filamentary structures connecting massive groups 
of galaxies in the SDSS. We perform a comparative analysis of some properties 
of galaxies in filaments, in the isotropic infall region, in the field, and in groups.
We study the luminosity functions (LF) and the dependence of the specific star 
formation rate (SSFR) on stellar mass, galaxy type, and projected distance to the groups
that define the filaments. 
We find that the LF of galaxies in filaments and in the isotropic infalling region
are basically indistinguishable between them, with the possible exception of late-type
galaxies. On the other hard, regardless of galaxy type, their LFs are clearly different 
from that of field or group galaxies. Both of them have characteristic absolute 
magnitudes and faint end slopes in between the field and group values.
More significant differences between galaxies in filaments and in the isotropic
infall region are observed when we analyse the SSFR. 
We find that galaxies in filaments have a systematically higher fraction of galaxies 
with low SSFR as a function of both, stellar mass and distance to the groups, indicating
a stronger quenching of the star formation in the filaments compared to both,
the isotropic infalling region, and the field.
Our results suggest that some physical mechanisms that determine the differences 
observed between field galaxies and galaxies in systems, affect galaxies even when they 
are not yet within the systems.
\end{abstract}
\section{Introduction} 
The large-scale structure of the universe is characterised by 
the presence of filaments which intersect at nodes wherein 
groups and clusters of galaxies are found (see \citealt{bond96}). 
Filaments are visually the most dominant structures in the distribution of 
galaxies and can be seen extending over scales up to tens of megaparsecs. 
The  hierarchical models of structure formation predict that groups and clusters 
grow by the continuous accretion of galaxies. This accretion usually happens 
along filaments in a non-isotropic way (e.g. \citealt{ebeling04}). 

Large galaxy redshift surveys such as the Sloan Digital Sky Survey 
(SDSS, \citealt{york00}) as well as N-body simulations have motivated the 
implementation of several methods to identify filaments 
{e.g. \citealt{stoica10,bond10,aragon10}). Many of the algorithms make use of 
the fact that filaments are the bridges that connect systems of galaxies 
\citep{pimbblet04,pimbblet05,colberg05,gonzalez10,smith12,zhang13,alpaslan14}. 
\citet{colberg05} suggested that the probability of finding a filament between
systems of galaxies is strongly dependent on their separation. 
\citet{zhang13} used the SDSS Data Release 8 \citep{dr8} and detected 
filaments using a sample of more than 50,000 clusters of galaxies. They 
selected cluster pairs separated by less than $35\mpc$ and found that richer 
clusters are connected to richer filaments.

Independently of the extension or the geometry, filaments are overdensities of 
galaxies and as such can affect the evolution of galaxies. The role of the 
filaments in this process has not been extensively studied. \citet{zhang13} 
studied the colour and luminosity distribution of galaxies in filaments. They found
that filament galaxies are bluer and fainter than cluster members. 
\citet{guo15} studied the satellite luminosity function of primary galaxies 
and found that the filamentary environment can increase by a factor of two the 
abundance of satellites compared with non-filament galaxies. They concluded 
that the filamentary environment may have a strong effect on the efficiency of
galaxy formation. 

\begin{figure*}
\includegraphics[width=140mm]{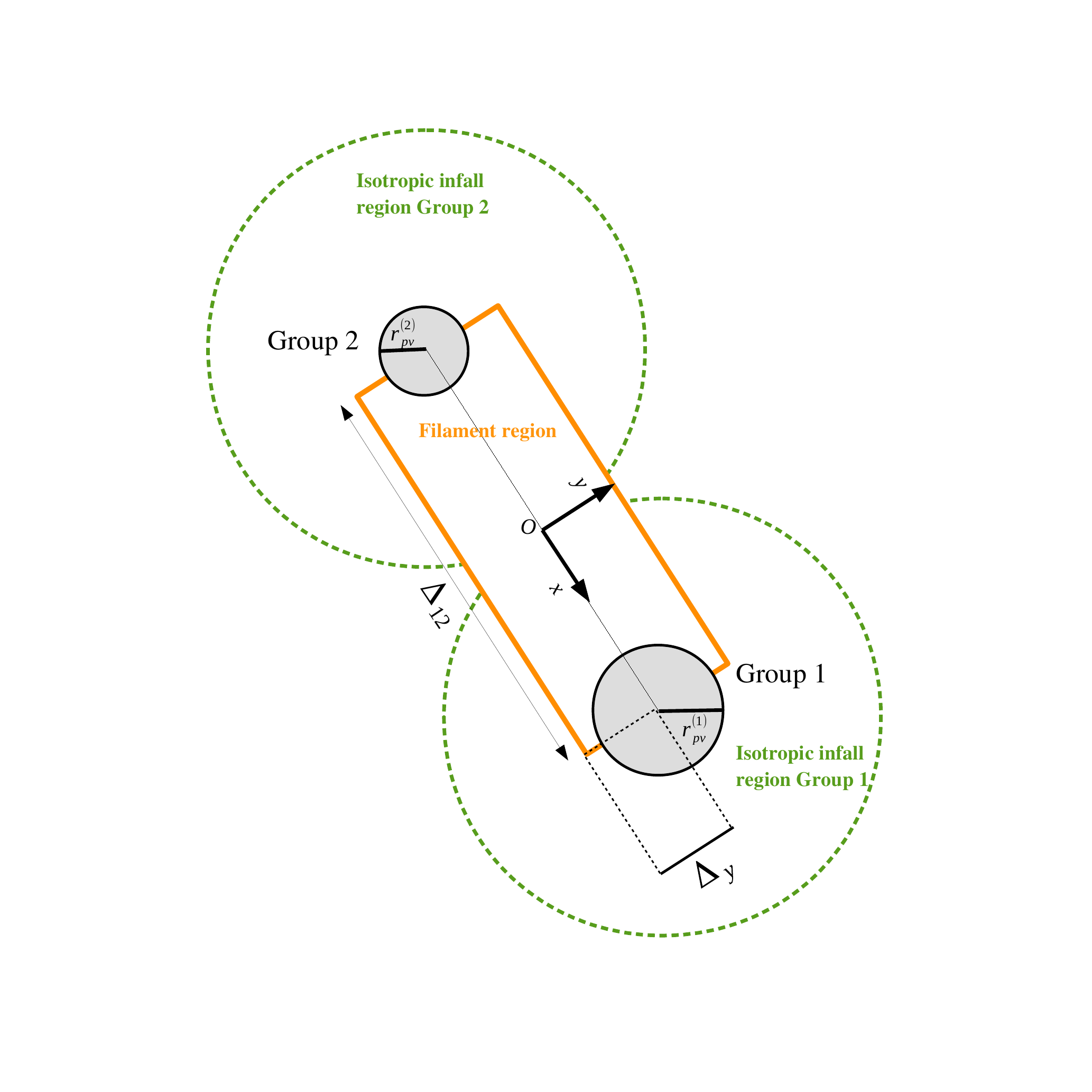}
\caption{A plane of the sky projection of two hypothetical galaxy groups 
(grey circles) and the geometry we use to define the infalling regions: green dashed circles
enclose the isotropic infall regions, and the orange rectangle defines the filament region. 
See text for details.
}
\label{geometry}
\end{figure*}

The region connecting filaments with clusters of galaxies is the so called 
infall region that extends from the outskirts of cluster up to several virial 
radii. \citet{porter08} found that galaxies falling into a cluster along 
filaments, are likely to undergo an enhancement of their star formation before 
they reach the virial radius of the cluster. Similarly, \citet{mahajan12} 
reported an excess of star forming galaxies in the outskirts of dynamically 
unrelaxed clusters and associated this phenomenon to the infall of galaxies 
through straight filaments. They concluded that a relatively high galaxy 
density in the infalling regions of unrelaxed clusters produced momentary 
bursts of star formation.

Regardless of the filamentary structure, the cluster infall region has been 
extensively studied. \citet{ellingson01} studied the composite radial 
distributions of different stellar populations as a function of clustercentric
radius. They found no evidence at any radius within the clusters for an 
excess of star formation over that seen in the field (see also 
\citealt{rines05} and \citealt{verdugo08}). The general agreement is that 
galaxy properties converge to those of field galaxies at $2-3$ virial radii. 
Group/filament preprocessing may play an important role in transforming 
galaxies before they enter into the cluster environment. It has also been 
suggested that a significant fraction of galaxies at large radii have passed 
through the core region of the cluster and have undergone environmental 
transformation within the virial radius (see \citealt{muriel14} and references
therein). 
It has also been studied the properties of galaxies in the infalling region at 
intermediate/high redshifts. \citet{just15} 
found evidences of preprocessing of galaxies in the infall region of clusters in 
the redshift range $0.4<z<0.8$. These authors found that at $z\sim0.6$, 
the fraction of red galaxies in the infall region is larger than in the field 
(see also \citealt{moran07} and \citealt{patel11})

In this paper, we study the population of galaxies in the infalling region of massive 
groups taken from \citet{ZM11}. We distinguish between galaxies infalling into groups
along filaments and those that are in the infalling region but outside filaments,
which we refer to as isotropic infalling galaxies. We aim to characterise how 
these two infalling regions affect the star formation in galaxies. 
Based on the fact that filaments are the bridges that connect systems of 
galaxies, we firstly implement an algorithm to search for filaments 
connecting pairs of massive groups. Then, we stack the galaxy population 
around these groups into a sample of galaxies in filaments, and another of
isotropically infalling galaxies. 
We compare these two populations with the galaxies in the groups
that are connected by the filaments, and with a sample of field galaxies.
This paper is organised as follows: we identify filaments using groups of galaxies 
and define our samples of galaxies in Sect. \ref{samples}; 
we compare the properties of galaxies in filaments and those
isotropically infalling into groups with field galaxies and galaxies in groups
in Sect. \ref{analysis}; we discuss our results and present our conclusions in 
Sect. \ref{discussion}. 
Throughout this paper we use Petrosian magnitudes, in the AB system, and 
corrected for Galactic extinction using the maps by \citet{sch98}. 
Absolute magnitudes and distances have been computed assuming a flat cosmological 
model with parameters $\Omega_0=0.3$, $\Omega_{\Lambda}=0.7$ and 
$H_0=100~h~{\rm km~s^{-1}~Mpc^{-1}}$. $K-$corrections have been computed
using the method of \citet{blantonk}~({\small KCORRECT} version 4.1). 
We have adopted a band shift to a redshift $0.1$ for the $r$ band 
(hereafter $^{0.1}r$), i.e. to approximately the mean redshift of the main 
galaxy sample of SDSS.

\begin{figure}
\includegraphics[width=90mm]{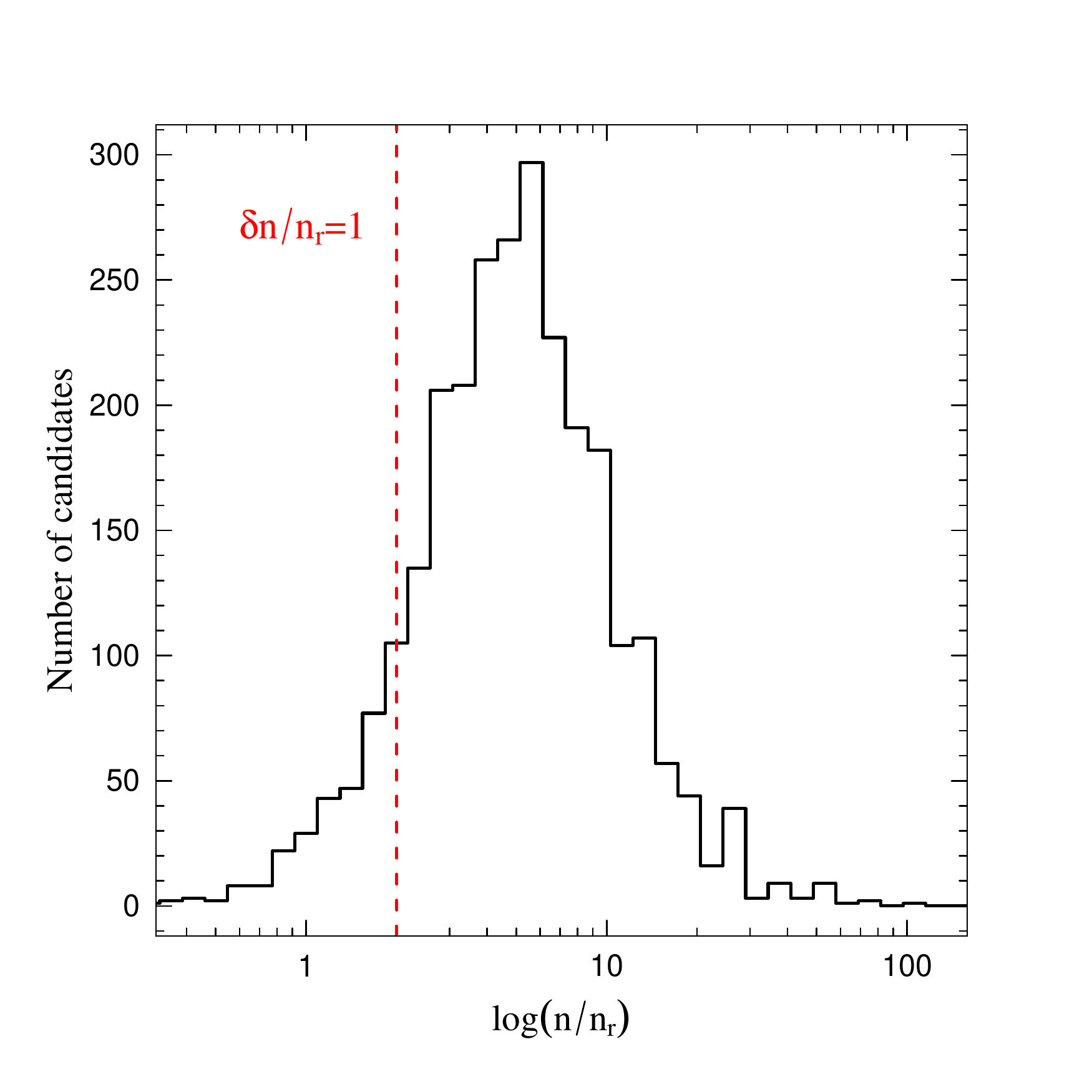}
\caption{The distribution of the galaxy number density (relative to the
random number density) in the filament region of our sample of group pairs. 
The vertical red dashed line corresponds to a galaxy
overdensity $\delta n/n_r=1$, which we use to define our sample of filaments.
}
\label{rho}
\end{figure}

\section{The samples}
\label{samples}
The purpose of this paper is to study how the star formation of galaxies
infalling into groups are differently affected by the environment, depending on 
whether they are in filaments or not.
The samples of groups used in this paper were drawn from the sample of groups 
identified by Zandivarez \& Mart\'inez 
(2011, hereafter ZM11) in the Main Galaxy Sample \citep{mgs} of the seventh 
data release of SDSS \citep{dr7}.
They used a standard friends-of-friends algorithm \citep{H&G:1982} to link galaxies into 
groups. The redshift-dependent linking length of the algorithm takes into 
account redshift space distortions. 
ZM11 implemented a complementary identification procedure using a higher 
density contrast in groups with at least 10 observed members, in order to split
merged systems and clean up spurious member detection. 
The authors computed group virial masses from the virial radius of the systems and the velocity 
dispersion of member galaxies \citep{Limber:1960,Beers:1990}. The catalogue of 
ZM11 comprises 15,961 groups with more than 4 members, adding up to 
103,342 galaxies. We refer the reader to \citet{ZM11} and references therein 
for further details of group identification. 

\subsection{Filaments connecting groups of galaxies}

A vast majority of the studies in the literature related to filaments, have focused 
the attention on filaments connecting clusters of galaxies. Since our aim
is not the creation of a complete catalogue of filaments connecting groups, 
we restrict our analysis to massive groups at this point. Arguably, this choice gives us
better chances of finding actual overdensities of galaxies stretching between systems.
From the ZM11 catalogue we select all groups with virial mass above the
catalogue's median mass ($\log(M_{\rm vir}/h^{-1}M_{\odot})\geq 13.5$)
and in the redshift range $0.05\leq z\leq0.15$. 
We use this subset to identify pairs (1,2) of groups defined by the 
following criteria: 1) the difference of the radial velocities of their 
baricentres ($\Delta V_{12}$) is less than a chosen value $\Delta V_{\rm max}$,
$|\Delta V_{12}|\leq \Delta V_{\rm max}$; 2) the projected distance between 
their baricentres ($\Delta_{12}$) is smaller than a given value 
$\Delta_{\rm max}$ while being two clearly separated groups in the sky, i.e., 
they are separated by a projected  distance larger than the sum of their projected virial 
radii: $r_{\rm pv}^{(1)}+r_{\rm pv}^{(2)}\leq \Delta_{12}\leq \Delta_{\rm max}$.
We choose $\Delta_{\rm max}=10\mpc$ and $\Delta V_{\rm max}=1000~\kms$. 
According to \citet{zandivarez03}, groups in the mass range under consideration here have 
redshift space correlation length $s_0\sim11\mpc$, thus, for the purposes of our work,
we do not search for inter-group filaments spanning larger redshift space distances. 
We call {\em nodes} to groups that are part of a pair according to 
the conditions 1) and 2). Nodes can be part of more than one pair.

Since filaments are overdense zones compared to the mean galaxy number density,
we select group pairs that are linked by overdensities in the galaxy 
distribution. We use all DR7 MGS galaxies in the redshift range under 
consideration and with apparent magnitudes $14.5 \le r \le 17.77$ and
proceed as follows:

\begin{enumerate}
\item Firstly, we clean up the MGS of all galaxies contained in cylinders
centred on groups and oriented along the line-of-sight with dimensions that 
escalate with group size. We find that we can exclude all galaxies in ZM11
groups if we choose the cylinders to have projected radius 
$1.7~ r_{\rm pv}$ and height $4.3~\sigma$, where 
$r_{\rm pv}$ and $\sigma$ are the projected virial radius 
and the velocity dispersion of the group, respectively.
This also excludes other galaxies that are geometrically close to the groups in 
redshift space, thus ours is a more conservative choice than 
only discarding galaxies in groups. 
\item We define a Cartesian coordinate system whose origin is located in the 
geometric centre ($O$) of each group pair with the $x-$axis oriented along the 
line connecting the centres of the groups, the $y-$axis orthogonal to the former 
in the plane of the sky, and the $z-$axis pointing outwards along the 
line-of-sight, see Fig.  \ref{geometry}. Hereafter, to avoid confusion with the 
letter we use to denote redshift, $z$, we will refer to the line-of-sight
axis as the $v-$axis.  
We consider the filament region to be a rectangular cuboid in redshift space
defined by: $|x|\leq \Delta_{12}/2$, $|y|\leq \Delta y$, and 
$|v|\leq \Delta V_{\rm max}$. We choose $\Delta y=1.5\mpc$. This size
is larger than the projected virial radii of $\sim99\%$ of the groups in our sample.
The plane-of-the-sky projection of this region is shown as an orange rectangle
in Fig. \ref{geometry}. 
\item We compute the galaxy overdensity in the filament region. To do so, we 
construct a random galaxy catalogue based on the MGS galaxies after excluding
group galaxies. The random catalogue is 100 times
denser, has the same redshift distribution and the same angular coverage than
the MGS galaxies. We further require that these random points do not lie within the
cylindrical volumes centred on groups that were used to clean up the MGS from galaxies 
in groups (see point (i) above). With this restriction, the volume filled by the random 
sample mimics that of the real data we use to identify filaments.
From all group pairs we select 
those having a number overdensity in the filament region 
$\delta n/n_r=(n-n_r)/n_r>1$, where $n$ and $n_r$ are the number of MGS galaxies 
and the normalised number of random points in the filament region, respectively. 
That is, we consider that a group pair is linked with a filament if the galaxy number
density in the filament region is at least twice the mean density at the pair's 
redshift. In Fig. \ref{rho} we show the distribution of $n/n_r$ of the group 
pairs, where the value $n/n_r=2$ corresponds to our choice for defining 
filaments. 
With this cut-off, out of our original sample of 3094 pairs, 2366
pairs hold filaments. It is worth noticing that most group pairs ($\sim76\%$) 
meet our overdensity condition and thus, most group pairs should be linked by 
actual filaments. 
\end{enumerate}

\begin{figure}
\includegraphics[width=90mm]{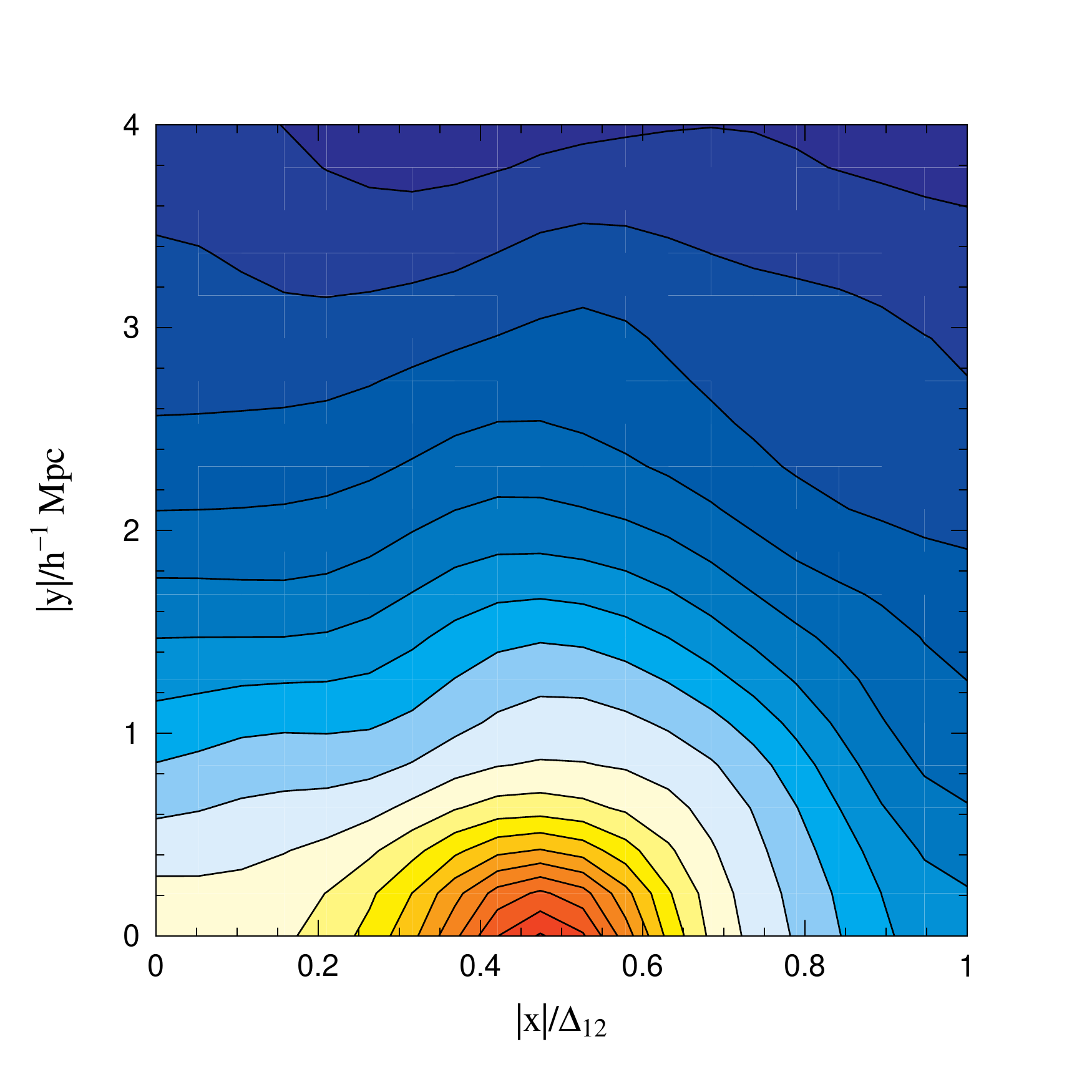}
\caption{The projected galaxy overdensity around the stacked sample of groups
with filaments. In these coordinate axes groups are centred at $(0.5, 0)$
and the filament region extends from there to the origin. See text for details. 
}
\label{overdensity}
\end{figure}

\begin{figure}
\includegraphics[width=90mm]{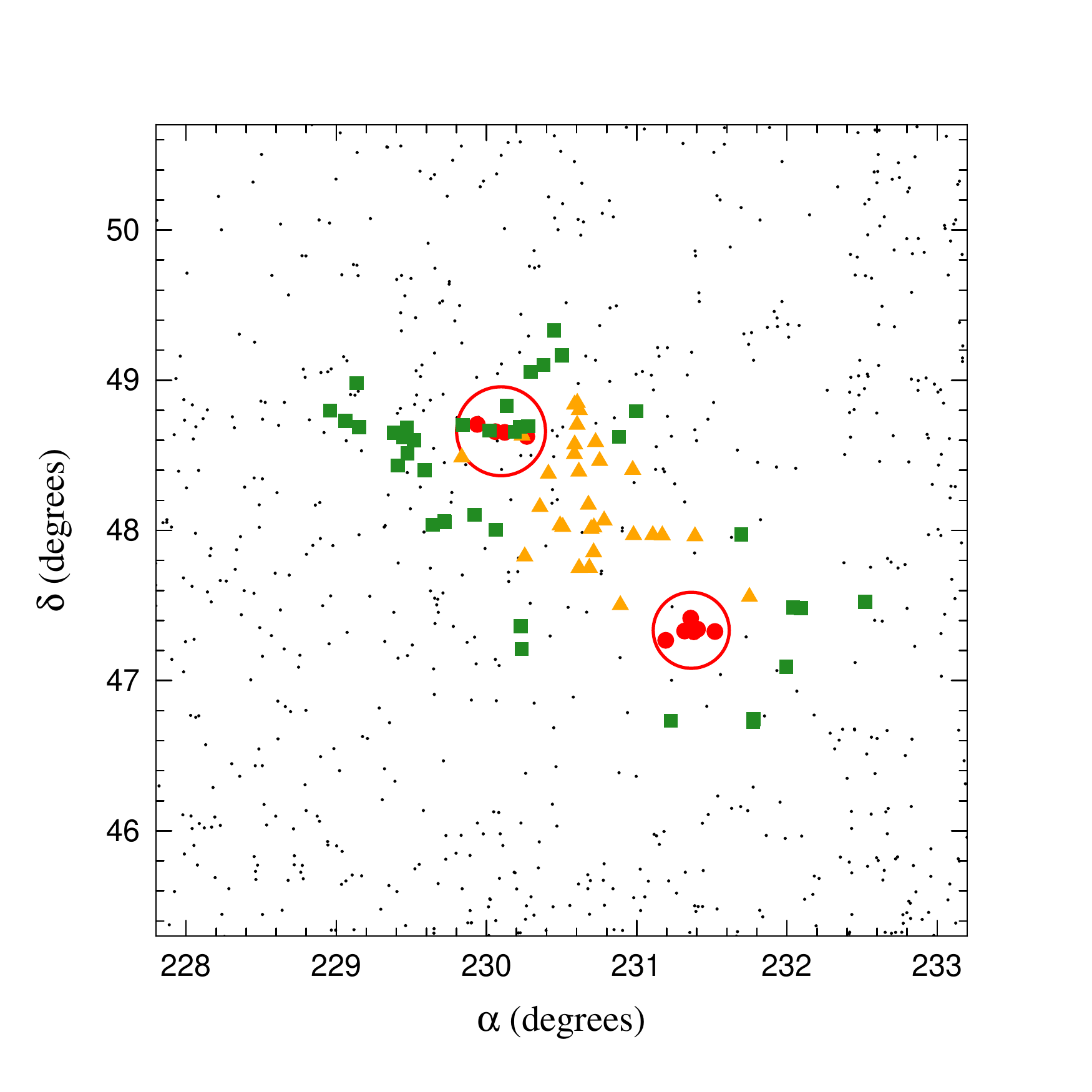}
\caption{An example of two groups of galaxies in our sample and the galaxies
we identify as infalling into them. Each group is marked by a circle which 
represents its projected virial radius. Galaxies belonging to each group are shown as
filled red circles. Isotropically infalling galaxies are shown as green
squares, while orange triangles represent galaxies in the filament region.
Field galaxies are shown as black dots.  
}
\label{example}
\end{figure}

\begin{figure}
\includegraphics[width=90mm]{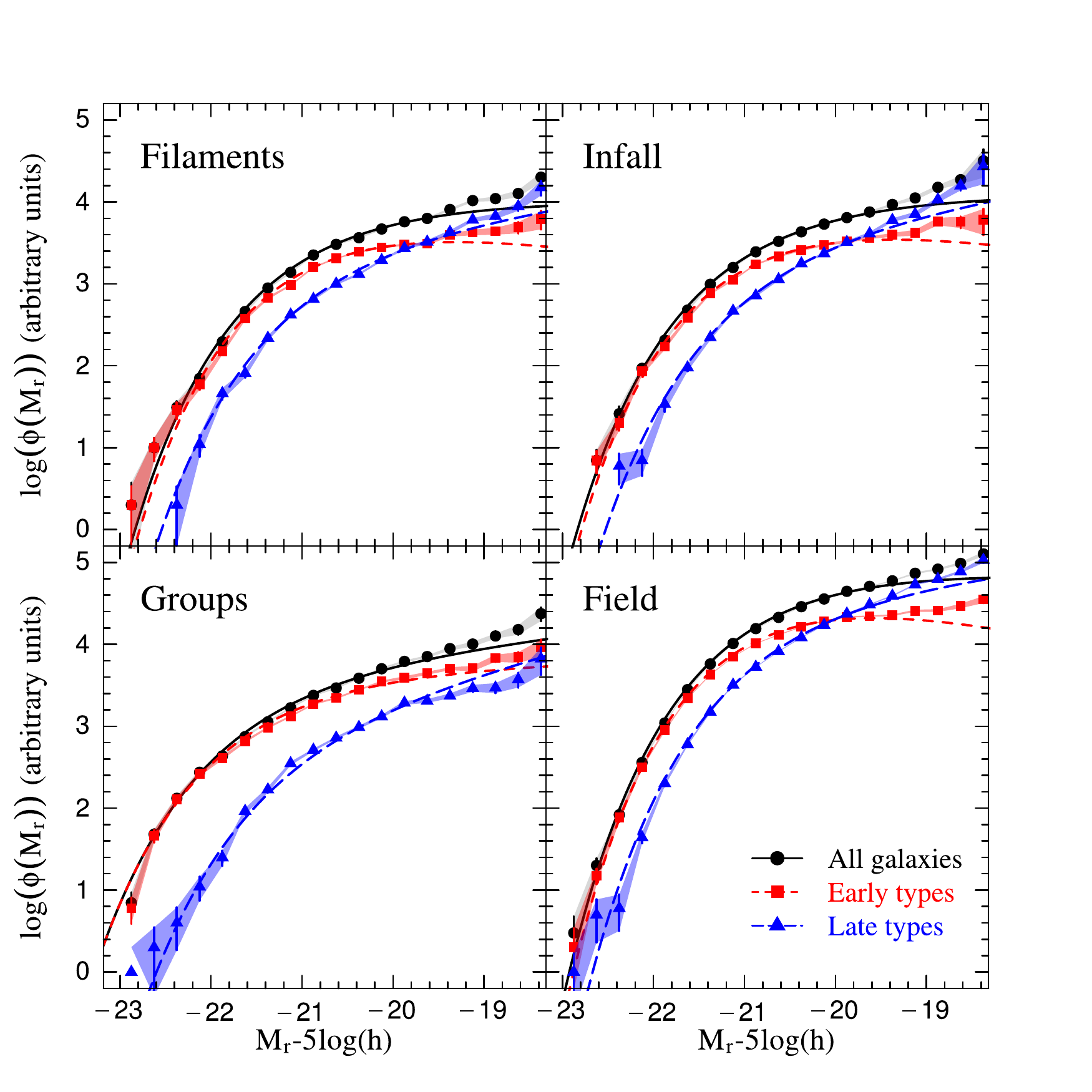}
\caption{The $^{0.1}r-$band luminosity functions of galaxies in the four 
different environments we probe: Filaments, isotropic infall area, groups and
field. We show in filled black circles the LF of all galaxies irrespective of 
their type; red squares correspond to early-type galaxies, and blue triangles
to late-type galaxies. 
Points were calculated using the $C^{-}$ method and error-bars using the
bootstrap resampling technique. Best fit Schechter functions were computed 
using the STY method. Best fit parameters are shown in Fig. \ref{sty}
below.
}
\label{lf}
\end{figure}

We show in Fig. \ref{overdensity} the galaxy overdensity in the plane of the sky
for our stacked sample of group pairs with filaments as a function of
$|x|/\Delta_{12}$ and $|y|$. In this figure, groups are located in the position
(0.5,0). Galaxies in groups have been removed. Besides the expected increase 
in density around the location of the groups, it is also clear the presence 
of an overdensity stretching from the groups towards the geometric centre of the 
group pairs, i.e., from $|x|/\Delta_{12}=0.5$ to $|x|/\Delta_{12}=0$. This contrasts
with the circular-like behaviour of the overdensity contours in the opposite 
direction.

\subsection{Infalling regions around groups: filaments and isotropic infall} 

We are particularly interested in exploring
possible differences between galaxies infalling along the preferred directions
defined by the filaments, and galaxies infalling from other directions, that we 
will consider to be infalling isotropically. 
Hereafter we will refer to the former as FG and the latter as IG.

For each group pair we consider the FGs as falling to/associated to
the closest group in projection. 
Every galaxy in the FG and IG samples will be considered as falling to/associated to
its closest group in projection. Thus, by construction, each group in a pair 
contributes to the sample of FG with galaxies that can be 
separated as far as $\sim \Delta_{12}/2$ in projection. We use that distance to define
the isotropic infall region around each group: a cylinder centred in the group
and oriented in the line of sight direction defined by a radius 
$\Delta_{12}/2$ and a height $2~\Delta V_{\rm max}$. We show the projection 
in the sky of these cylinders as green dashed circles in Fig. \ref{geometry}.
Each group contributes to the IG sample up to the same scale as it does to the 
sample of galaxies in filaments. An example of and actual group pair 
along with the galaxies in its filament region and in the isotropic regions, can 
be seen in Fig. \ref{example}.
The samples of IG and FG comprise 33,094 and 26,043 galaxies, respectively.
According to the criteria described above, a galaxy can not be classified simultaneously
as both: FG and IG. Nevertheless, it is worth mentioning that a FG can be a member of more 
than one filament.

Our samples of IG and FG include galaxies that are effectively in the 
isotropic infalling region or in the filaments, respectively. 
It is clear that, both samples will be contaminated by 
foreground and background galaxies, and backsplash galaxies.
Unless we had three dimensional positions and 
velocities, we are unable to isolate the actual IG and FG samples. 
By construction, the IG and FG samples have similar redshift distributions, 
and we expect both to be contaminated with foreground and background in the same way. 
Therefore, any difference in the galaxy properties of the samples should reflect an 
actual and more significant difference in the populations.

\begin{figure}
\includegraphics[width=80mm]{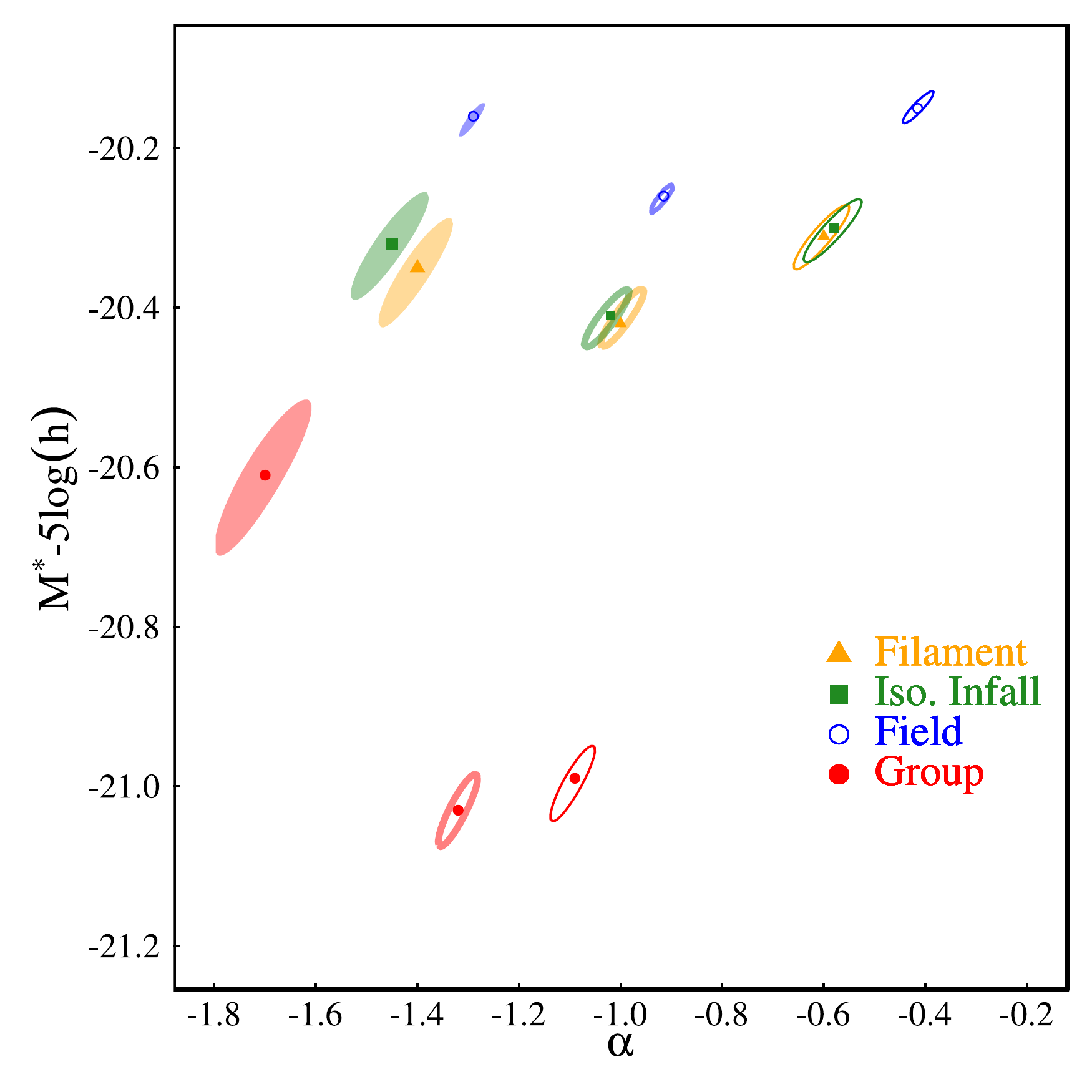}
\caption{Best fit Schechter parameters of the luminosity functions shown in Fig. 
\ref{lf} and quoted in Table \ref{table_sty}. 
Points are the best fitting values, shown along their $1~\sigma$
contours. Different colours and symbols indicate the environment: blue open circles correspond 
to field galaxies; green squares to IG, orange triangles to FG, and red filled circles to galaxies in groups.
Different type of contours indicate galaxy type: filled contours correspond to late types, contours enclosed
by thin lines correspond to early types, and contours enclosed by thick lines
correspond to all galaxies, irrespective of their type. 
}
\label{sty}
\end{figure}

\subsection{Control samples: field and group galaxies}

To understand the effects of the infall regions on galaxies, an adequate comparison
with samples of galaxies in the field and in groups is needed.

We construct a sample of field galaxies drawn from the MGS DR7 by randomly 
selecting galaxies in the redshift range under consideration, avoiding in the process
all MZ11 groups, filaments and isotropic infall regions. We impose to this sample of
field galaxies to have the same redshift distribution as the 
FG and IG samples. Our resulting sample of field galaxies comprises 156,357 galaxies.

The sample of galaxies in groups contains all galaxies in the groups with filaments,
adding up a total of 19,464  galaxies.

\section{Comparing galaxy populations in the infalling regions, field and groups}
\label{analysis}
In this section we perform a comparison of properties of the galaxy population
inhabiting filaments, isotropic infall region, field, and groups. We focus our analyses
on the luminosity and the star formation of galaxies.

\subsection{Luminosity function}

\begin{table}
\caption{Best fitting Schechter's parameters of the luminosity functions shown
in Fig. \ref{lf} computed through the STY method. See also Fig. \ref{sty}.}  
\begin{tabular}{llcc}
\hline
Environment & Galaxy type & $\alpha$ & $M^{\star}-5\log(h)$   \\
            &             &          &  $^{0.1}r-$band        \\
\hline
Field     & All        & $-0.91\pm0.02$ & $-20.26\pm0.02$ \\
          & Early type & $-0.41\pm0.02$ & $-20.15\pm0.01$ \\
          & Late type  & $-1.29\pm0.02$ & $-20.16\pm0.02$ \\
\hline
Isotropic & All        & $-1.02\pm0.05$ & $-20.41\pm0.04$ \\
Infall    & Early type & $-0.58\pm0.06$ & $-20.30\pm0.04$ \\
          & Late type  & $-1.45\pm0.07$ & $-20.32\pm0.07$ \\
\hline
Filament  & All        & $-1.00\pm0.05$ & $-20.42\pm0.04$ \\
          & Early type & $-0.60\pm0.06$ & $-20.31\pm0.04$ \\
          & Late type  & $-1.40\pm0.07$ & $-20.35\pm0.07$ \\
\hline
Groups    & All        & $-1.32\pm0.04$ & $-21.03\pm0.05$ \\
          & Early type & $-1.09\pm0.04$ & $-20.99\pm0.05$ \\
          & Late type  & $-1.70\pm0.09$ & $-20.6\pm0.1$   \\
\hline
\label{table_sty}
\end{tabular}
\end{table}

We use two methods to compute the $^{0.1}r-$band LF of galaxies in our samples: the 
non-parametric $C^{-}$ \citep{lb71,cho87} for the binned LF, and the STY method 
\citep{sty} to compute the best-fit \citet{schechter} function parameters:
the faint-end slope $\alpha$, and the characteristic absolute magnitude $M^{\ast}$.
We also compute separately the LF of early and late type galaxies according to their
concentration parameter \citet{Strateva:2001}.
In Fig. \ref{lf} we show the resulting binned LFs in arbitrary units, also shown
in this figure are the best fitting Schechter functions. The Schechter parameters 
along with their $1\sigma$ error contours are shown in Fig. \ref{sty} separately for
the complete samples and the subsamples of early and late-types. 

It is easier to spot similarities and differences between the populations
by inspecting Fig. \ref{sty}. As expected, and regardless of type, field and group 
galaxies are the two extremes cases: the former have the shallowest faint-end slope 
and the faintest characteristic absolute magnitude, while the latter are the opposite.
In between them, the LF parameters of FGs and IGs are closer to, however
different from, those of field galaxies.
The LF of FG and IG are indistinguishable for the complete sample of galaxies 
and for early-types. A subtle difference is observed for late-types: FG have a 
shallower faint end slope, and their characteristic magnitudes is brighter. This is,
however, only an one sigma difference.
It is worth noticing that the values of $\alpha$ and $M^{\ast}$ for our sample 
of galaxies in groups, regardless of galaxy type, are fully consistent with 
the results by ZM11 for groups with masses in the range of our sample.
A straightforward comparison for field galaxies can not be made with previous
determinations of the LF of galaxies (e.g. \citealt{montero09}),
given our particular definition of field galaxies, which excludes galaxies in 
groups and in the infalling regions of groups.

\begin{figure*}
\includegraphics[width=160mm]{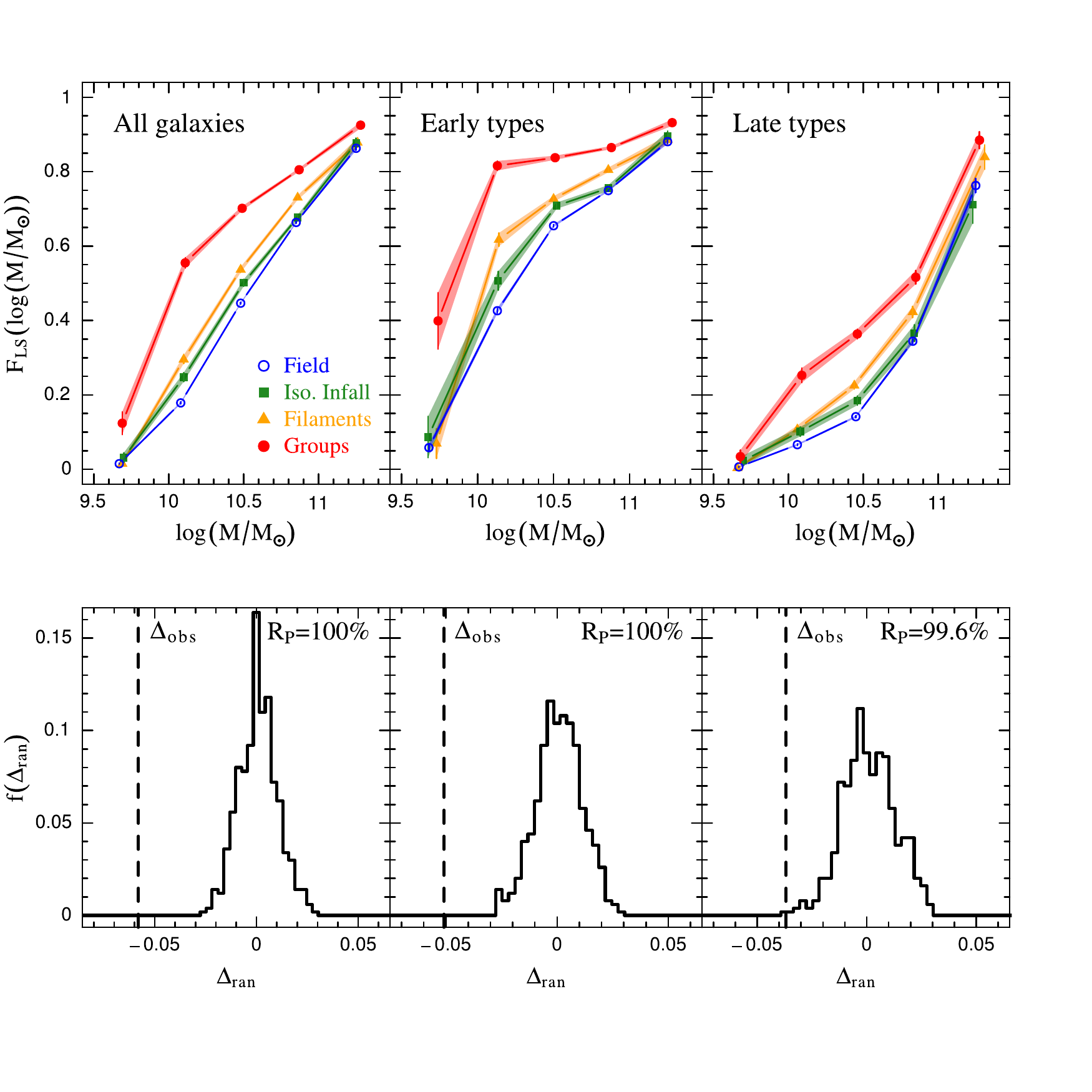}
\caption{The dependence of the SSFR on stellar mass. 
{\em Upper panels}: the fraction of low SSFR ($\log({\rm SSFR/yr^{-1}})<-11$)
galaxies, $F_{\rm LS}$, as a function of stellar mass. 
{\em Left panel} includes all galaxies,
while the {\em centre} and {\em right} panels consider only early and late types, 
respectively. Errorbars were computed using the bootstrap resampling technique.
In the {\em lower panels}, we show the results of applying the test
of section \ref{SSFR} for the null hypothesis that there are no differences in 
the SSFR as a function of stellar mass between the IG and FG samples of the 
{\em upper panels}. Each panel shows the normalised distributions of the quantity 
$\Delta_{\rm ran}$, the values of $\Delta_{\rm obs}$ (vertical dashed lines), 
and the rejection probability,
$R_P$, of the null hypothesis (see table \ref{table_ssfr_mass}). 
}
\label{fraction_ssfr_mass}
\end{figure*}
\subsection{Specific star formation rate}
\label{SSFR}
We search now for differences in the star formation of the galaxies in our samples.
In particular, we focus our attention on the specific star formation rate,
and its dependence with stellar mass and the distance to the nodes of the filaments.

The values of stellar mass and SSFR for the galaxies
in our samples have been extracted from the MPA-JHU DR7 release of spectra 
measurements\footnote{\tt http://www.mpa-garching.mpg.de/SDSS/DR7}. This catalogue
provides, among other parameters, stellar masses based on fits to the photometry 
following \citet{kauffmann03} and \citet{salim07}, and star formation rates based 
on \citet{brinchmann04}.

In Fig. \ref{fraction_ssfr_mass} we show the fraction of low SSFR 
($\log({\rm SSFR/yr^{-1}})<-11$) as a function of stellar mass for our samples 
of galaxies. Lowest mass bins are numerically dominated by late-types and 
the highest mass bins by early-types. 
Regardless of whether we consider all galaxies, or whether we split 
them into early and late types, group and field galaxies exhibit the extreme values: 
over the stellar mass range we probe, groups have the highest fraction of low SSFR 
galaxies, while the opposite occurs for field galaxies.
In between them, but typically closer to field values are the median SSFR of FG and IG.
Filaments have systematically a fraction of low SSFR galaxies higher than the isotropic 
infall region, for stellar masses higher than $\sim 10^{10} M_{\odot}$.

To check whether these differences between FG and IG are indicative of actual 
differences between the populations we rely on the test used by \citet{muriel14}. 
Briefly, let us consider two samples of objects, $A$ and $B$, and two physical
quantities $X$ and $Y$. This test allows us to tell whether
the two populations have different trends of $Y$ as a function of $X$.
Let us consider $N_{\rm bin}$ bins of the $X$ variable, and let $N_A^{(i)}$ and 
$N_B^{(i)}$ be the numbers of objects of the samples $A$ and $B$ in the $i-$th bin, 
respectively.
Now let us consider, within each bin, the sum of the differences in the quantity $Y$ 
from all the pairs formed by one object from the sample $A$ and the other from the 
sample $B$. After accumulating over all bins and normalising by the number of pairs
used in the process, we arrive to the quantity:
\begin{equation}  
\Delta_{\rm obs}\equiv\frac{\sum_{i=1}^{N_{\rm bin}}\sum_{j=1}^{N_A^{(i)}}
\sum_{k=1}^{N_B^{(i)}}
\left(Y^{(i)}_{A,j}-Y^{(i)}_{B,k}\right)}
{\sum_{l=1}^{N_{\rm bin}}N_A^{(l)}N_B^{(l)}},
\label{delta_obs}
\end{equation}
where $Y^{(i)}_{A,j}$ and $Y^{(i)}_{B,k}$ are the $Y$ values of the $j-$th 
object of the sample $A$ and the $k-$th object of the sample $B$ in the $i-$th bin,
respectively. With this definition, $\Delta_{\rm obs}$ is a measure of the average
difference in the quantity $Y$ between the two samples, once the overall trend of $Y$ as a 
function of $X$ has been removed.
If the two samples had no differences regarding the behaviour of $Y$ as a
function of $X$, then we would expect $\Delta_{\rm obs}\simeq 0$. On the other hand, 
a non-zero value can not be straightforwardly interpreted as mirroring a significant 
difference between the samples, unless the obtained value is unlikely
for the null hypothesis corresponding to the case in which both samples are drawn 
from the same underlying population.   

Let $C$ be the sample resulting by the merging of samples $A$ and $B$. By construction,
this sample has $N_C^{(i)}=N_A^{(i)}+N_B^{(i)}$ in the $i-$th bin.
Now, let us randomly select from $C$ two subsamples $A'$ and $B'$, bound to have
the same number of objects per bin as the samples $A$ and $B$, respectively. 
Clearly, each of these new samples will include objects from both, $A$ and $B$.
From the samples $A'$ and $B'$ we compute the value $\Delta_{\rm ran}$.
If we repeat this procedure a large number of times, performing a different random
selection each time,  $\Delta_{\rm ran}$ will be distributed around the value $0$. 
Provided $\Delta_{\rm obs}>0$, we can now quantify the rejection probability of the 
null hypothesis by computing the fraction, $F$, of random realisations that give 
$\Delta_{\rm ran}>\Delta_{\rm obs}$: the rejection probability will be $R_P=1-F$. 
In the case $\Delta_{\rm obs}<0$, $F$ is defined as the fraction of random
realisations that give $\Delta_{\rm ran}<\Delta_{\rm obs}$. The sign of 
$\Delta_{\rm obs}$ tell us which of the samples has systematically larger
values of the quantity $Y$ as a function of $X$.

Results of 1000 random realisations of the test can be seen in the bottom panels
of Fig. \ref{fraction_ssfr_mass}, where we use $X=\log(M)$ and $Y=$ SSFR.
The test is conclusive: irrespective of galaxy type, the null hypothesis of 
FG and IG being drawn from the same underlying population is ruled out at a
confidence level above $99\%$ in the three cases. 
This means that FG and IG have significant different trends in their SSFR 
as a function of stellar mass.
Besides the surrounding large scale structure, galaxies can be affected by their 
small-scale local environment. Evidence has been found that galaxies in pairs have
their star formation suppressed by each other, see for instance \citet{alpaslan15}.
To test whether our results could be due to a relative excess of pairs in the FG sample, 
we compute the fraction of galaxies in the FG and IG samples that are in pairs. 
For each galaxy in these samples, we search for a pair in the MGS, using the criterion 
by \citet{alpaslan15}. We find that there are no significant differences between the 
fraction of galaxies in pairs in FG (7.3\%) and in IG (7.1\%).
\begin{table}
\caption{The dependence of SSFR on the stellar mass:
results of applying the test of section \ref{SSFR}
to the samples of galaxies in the isotropic infall region and the sample
of galaxies in filaments. See also Fig. \ref{fraction_ssfr_mass}.}
\begin{tabular}{lcc}
\hline
Galaxy type & $\Delta_{\rm obs}$               & Rejection   \\
            & ($\log(\rm{SSFR}/\rm{yr}^{-1})$) & probability \\
\hline
All         & $0.079$ &  100\% \\
Early types & $0.069$ &  100\% \\
Late types  & $0.040$ &  99.9\% \\
\hline
\label{table_ssfr_mass}
\end{tabular}
\end{table}

\begin{table}
\caption{The dependence of SSFR on the projected distance to the
nodes. Results of applying the significance test of section \ref{SSFR}
to the samples of galaxies in filaments and in the isotropic infall region (Fig. 
\ref{fraction_ssfr_dist}).}
\begin{tabular}{lccc}
\hline
Galaxy type  & Stellar mass range   & $\Delta_{\rm obs}$ & Rejection \\
        & ($\log(M/M_{\odot})$)& ($\log(\rm{SSFR}/\rm{yr}^{-1})$) & probability \\
\hline
All         & $9.40-10.13$  & $0.157$ &  96.8\% \\
Early types & $9.40-10.13$  & $0.728$ &  93.6\% \\
Late types  & $9.40-10.13$  & $0.019$ &  86.0\% \\
\hline
All         & $10.13-10.87$ & $0.055$ &  100\%  \\
Early types & $10.13-10.87$ & $0.041$ &  99.8\% \\
Late types  & $10.13-10.87$ & $0.035$ &  99.4\% \\
\hline
All         & $10.87-11.60$ & $0.047$ &  100\%  \\
Early types & $10.87-11.60$ & $0.033$ &  99.4\% \\
Late types  & $10.87-11.60$ & $0.133$ &  100\%  \\
\hline
\label{table_ssfr_dist}
\end{tabular}
\end{table}
\begin{figure*}
\includegraphics[width=160mm]{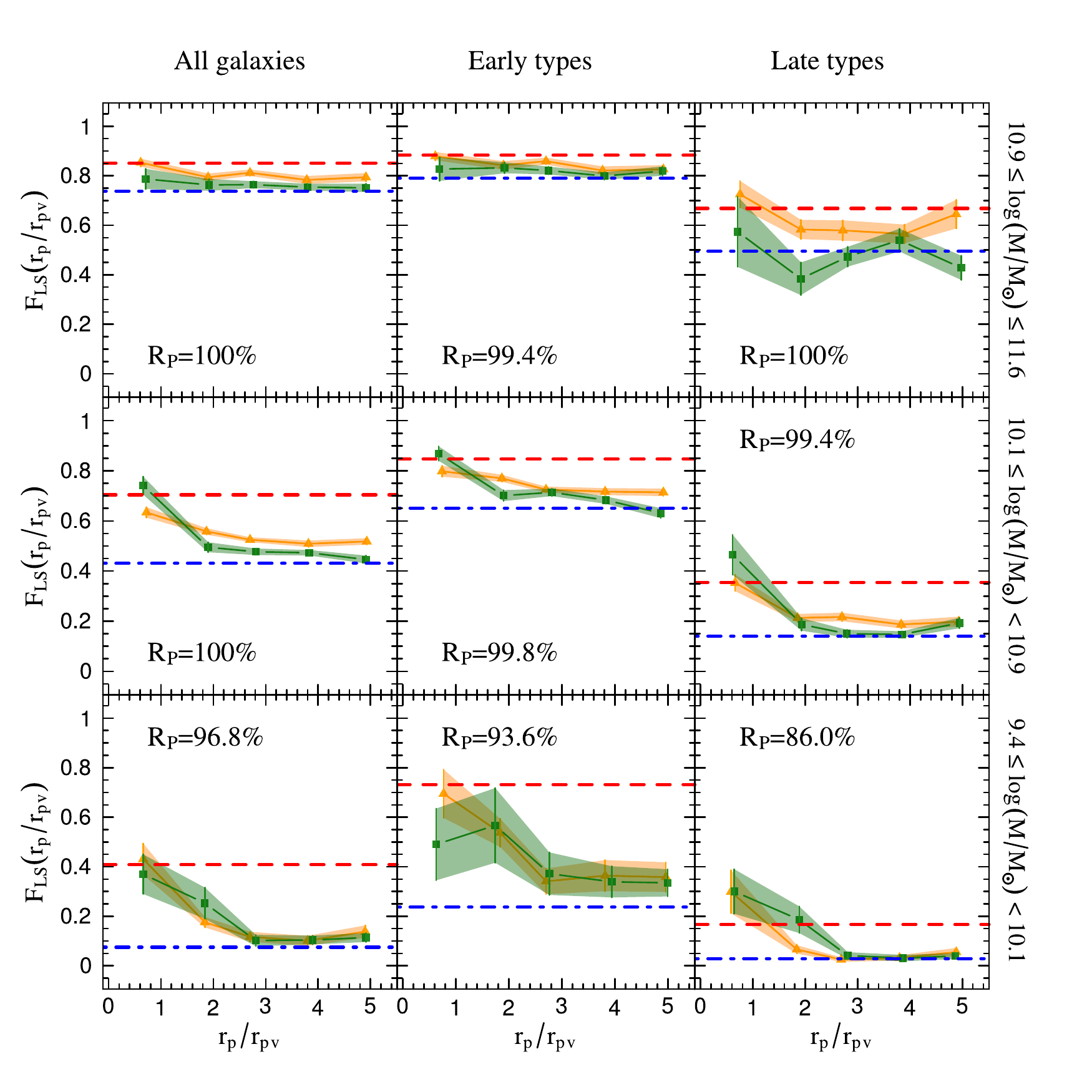}
\caption{
The fraction of low SSFR ($\log({\rm SSFR/yr^{-1}})<-11$) galaxies, $F_{\rm LS}$,
as a function of the projected distance to the nearest node, $r_p$, in units of the 
projected virial radius of the nearest node, $r_{\rm pv}$.  
{\em Left column}: all galaxies, {\em central column}: early-type galaxies, and
{\em right column}: late-type galaxies. Rows correspond to the
mass intervals quoted in the right side of the figure. Galaxy mass decreases 
from top to bottom. Green squares represent IG and orange
triangles FG. Errorbars were computed using the bootstrap resampling technique.
Horizontal red long-dashed lines, and blue dashed-dotted lines are the fraction of 
low SSFR galaxies in groups and in the field, respectively.
We quote inside each panel the rejection probability, $R_P$, of the null 
hypothesis that there are no differences in the SSFR as a function of 
projected distance between the IG and FG samples (see table \ref{table_ssfr_dist}).  
}
\label{fraction_ssfr_dist}
\end{figure*}

Our results so far, show that both, FG and IG are different from field galaxies
(and group galaxies) in terms of their SSFR. For a galaxy, the fact of
being close to a massive group modifies its star formation rate. Even more, 
how a galaxy's star formation is affected depends on whether it is located in a 
filament or in the isotropic infall zone. 
In what follows, we analyse how the 
impact of the environment on star formation depends on distance to the nodes.

In Fig. \ref{fraction_ssfr_dist} we show the fraction of low SSFR galaxies, 
as a function of the projected distance to the nearest node in units of the node's 
projected virial radius. Since SSFR depends on stellar mass,
we split our samples into three bins of stellar mass 
(rows in Fig. \ref{fraction_ssfr_dist}),
and also analyse separately each galaxy type (columns in Fig. 
\ref{fraction_ssfr_dist}).
We also show in this figure, the mean values corresponding to field and group galaxies
as dashed horizontal lines. 
As a general trend, the fraction of low SSFR galaxies rises towards the nodes, and
smoothly decreases outwards. We find that:
\begin{enumerate}
\item For massive galaxies, which are dominated in number by early-types,
the dependence of the fraction of low SSFR on distance, for both FG and IG,
is quite flat. 
FG have a higher fraction of low SSFR galaxies over the whole range.
The fraction of low SSFR IG is either, consistent with, or marginally
larger than the field value. In any case, no differences with the field are observed
beyond $\sim 3~ r_{\rm vir}$. On the other hand, FG do not reach field values,
even though their low SSFR fraction decreases with distance.
The rejection probability in all cases indicates that the differences between IG and FG 
are significant. For late types IG, it is interesting to note the data point at 
$r_{\rm p}\sim 2~r_{\rm vir}$, which is a $\sim 2~\sigma$ signal indication of an 
enhancement in the star formation. A similar effect has been reported for galaxies 
infalling into clusters by \citet{porter08} and \citet{mahajan12}.  
\item For intermediate mass galaxies, the effects of the filamentary environment on 
the star formation are stronger. FG do not reach field values in the whole range probed.
IG are marginally more affected in scales $r_p \leq 1.4 r_{\rm vir}$.
Again, the rejection probability in all cases means that differences between IG and FG 
are significant.
\item For low mass galaxies, numerically dominated by late-types, no significant 
differences between FG and IG are seen, this is mirrored by the low values
of the rejection probability. No significant departures from field values are observed
beyond $\sim 3 r_{\rm vir}$ for the whole sample and for late types. However, for 
early types, field values are not reached at all over the whole range explored. 
\end{enumerate}

\section{Discussion and conclusions}
\label{discussion}

In this paper we study the effect of environment upon galaxies infalling into groups.
For this purpose, we search for filamentary structures connecting massive groups of 
galaxies using samples of groups and galaxies taken from the SDSS DR7. 
We compare properties of galaxies around these filaments' nodes, selected according 
to their projected distances and their radial velocity difference. We classify 
galaxies in the vicinity of these groups in two cases: those that are in the filament 
region (FG), and those that are in the isotropic infall region (IG). It is clear that 
both of these samples suffer from contamination from foreground and background 
galaxies. The use of control samples of field, and group galaxies, allows us to spot 
the actual differences between infalling galaxies and galaxies in the field and in 
groups.

Our comparison focuses in two physical properties of galaxies: luminosity, by means of 
the analysis of the luminosity functions; and, specific star formation rate, 
by studying its dependence on stellar mass, galaxy type, and projected distance to the 
nodes.
 
We find that the luminosity functions of FG and IG galaxies are basically 
indistinguishable between them, with the possible exception of late-types. 
On the other hand, and regardless of type, both of them are clearly different from 
those of field or group galaxies. Galaxies in filaments and in the isotropic infall 
region have characteristic absolute magnitudes brighter by $\sim0.2$ magnitudes when 
compared to field galaxies, and fainter by $\sim0.6$ magnitudes compared to galaxies 
in the nodes. These differences are 
larger when we consider early types, and smaller when we consider late types.
A similar effect is seen in the faint end slope: it is larger (in absolute values) 
in $\sim 0.1$ for LG and IG when compared to the field, and smaller in $\sim 0.3$
when compared to the nodes' value. Again, differences are larger for early types, and
smaller for late types. Due to the apparent magnitude limit of the main galaxy sample 
of SDSS, we are not probing faint magnitudes, and thus, the faint end slope of the 
LF is basically computed from the convexity of the LF around the characteristic 
magnitude. It is clear that, regarding the luminosity, FG and IG differ from both: 
field galaxies and galaxies in the nodes. 

Significant differences between FG and IG appear when we analyse the SSFR. 
Regardless of type, all samples analysed here, have different trends of SSFR as
a function of stellar mass. The samples of field and group galaxies are the two 
extremes, with the lowest and highest fraction of low SSFR galaxies, respectively. 
In between them, FG have a significantly larger fraction of low SSFR galaxies than IG.
Thus, not only these two samples differ from the field and the groups, but also they
have been affected differently by the environment. Clearly, galaxies infalling
into groups along filamentary structures have experienced a stronger quenching in 
their star formation than galaxies infalling into groups from other directions. 

Another clear indication that filamentary structures have a distinct impact on galaxy
evolution appears when we analyse the dependence of the SSFR with the projected 
distance to the nodes of the filaments. We find that the fraction of low SSFR 
galaxies increases towards the nodes and decreases outwards for both, IG and FG. 
However, FG have a typically larger fraction of low SSFR galaxies compared to IG up to 
$\sim 3 r_{\rm vir}$ projected radii. These differences is better seen for high and
intermediate mass galaxies. Our results also show that filaments affect star formation
further out than the isotropic infalling region does. 
Over the range of distances we probe, we do not find any enhancement of the
star formation relative to the field, with the exception of massive late types IG.
These galaxies have an excess of high SSFR galaxies compared to the field at a
projected distance of $r_{\rm p}\sim 2~r_{\rm vir}$.
This may be consistent with the findings of \citet{porter08} and \citet{mahajan12}
in the outskirts of clusters. 

Our results show that galaxies infalling into massive groups differ from field galaxies
regarding their star formation up to $\sim 3 r_{\rm vir}$, and even further out if 
they are located in filaments. Galaxies infalling into groups have lower star 
formation than field galaxies, but still not as low as group galaxies. This quenching 
of star formation is stronger in filaments. On the other hand, both, FG and IG are 
already typically brighter than field counterparts, 
still not as much as group galaxies. These two results are an indication that
some of the physical mechanisms that determine the differences observed between field
galaxies and galaxies in systems, affect galaxies even when they are not yet within the
systems.

\section*{Acknowledgements}
We thank the referee, Mehmet Alpaslan, for his comments which improved the paper.
This work was supported with grants from CONICET
(PIP 11220120100492CO and 11220130100365CO) and SECYT-UNC, Argentina.
Funding for the Sloan Digital Sky Survey (SDSS) has been provided by the 
Alfred P. Sloan 
Foundation, the Participating Institutions, the National Aeronautics and Space 
Administration, the National Science Foundation, the U.S. Department of Energy, 
the Japanese Monbukagakusho, and the Max Planck Society. The SDSS Web site is 
http://www.sdss.org/.
The SDSS is managed by the Astrophysical Research Consortium (ARC) for the 
Participating Institutions. The Participating Institutions are The University 
of Chicago, Fermilab, the Institute for Advanced Study, the Japan Participation 
Group, The Johns Hopkins University, the Korean Scientist Group, Los Alamos 
National Laboratory, the Max Planck Institut f\"ur Astronomie (MPIA), the 
Max Planck Institut f\"ur Astrophysik (MPA), New Mexico State University, 
University of Pittsburgh, University of Portsmouth, Princeton University, 
the United States Naval Observatory, and the University of Washington.


\label{lastpage}
\end{document}